\documentclass[a4paper,fleqn,usenatbib]{mnras}

\usepackage[T1]{fontenc}
\usepackage{ae,aecompl}

\usepackage{graphicx}	% Including figure files
\usepackage{amsmath}	% Advanced maths commands
\usepackage{amssymb}	% Extra maths symbols

\def\msun{\hbox{M$_\odot$}}

\title[New LMC star cluster]{Discovery of a loose star cluster in the Large Magellanic Cloud}

\author[A.E. Piatti]{
Andr\'es E. Piatti$^{1,2}$\thanks{E-mail: andres@oac.uncor.edu (AEP)}
\\
$^{1}$Observatorio Astron\'omico, Universidad Nacional de C\'ordoba, Laprida 854, 5000, 
C\'ordoba, Argentina\\
$^{2}$Consejo Nacional de Investigaciones Cient\'{\i}ficas y T\'ecnicas, Av. Rivadavia 1917, 
C1033AAJ, Buenos Aires, Argentina\\
}

\date{Accepted XXX. Received YYY; in original form ZZZ}

\pubyear{2015}

\begin{document}
\label{firstpage}
\pagerange{\pageref{firstpage}--\pageref{lastpage}}
\maketitle

\begin{abstract}
We present results for an up-to-date uncatalogued star cluster projected
towards the Eastern side of the Large Magellanic Cloud (LMC) outer disc.
The new object was discovered from a search of loose star cluster in the
Magellanic Clouds' (MCs) outskirts using kernel density estimators on Washington
$CT_1$ deep images. Contrarily to what would be commonly expected, the
star cluster resulted to be a young object (log($t$ yr$^{-1}$) = 8.45) with
a slightly subsolar metal content ($Z$ = 0.013) and a total mass of 650$\msun$.
Its core, half-mass and tidal radii also are within the frequent values of
LMC star clusters.
However, the new star cluster is placed at the Small Magellanic Cloud distance 
and at 11.3 kpc from the LMC centre. We speculate with the possibility that
it was born in the inner body of the LMC and soon after expeled into the 
intergalactic space during the recent Milky Way/MCs interaction.
Nevertheless,
radial velocity and chemical
abundance measurements are needed to further understand its origin, as well as
extensive search for loose star cluster in order to constrain the effectiveness of
star cluster scattering during galaxy interactions.

\end{abstract}


%%%%%%%%%%%%%%%%%%%%%%%%%%%%%%%%%%%%%%%%%%%%%%%%%%

%%%%%%%%%%%%%%%%% BODY OF PAPER %%%%%%%%%%%%%%%%%%

\section{Introduction}

Star clusters in the outer disc of the Large Magellanic Cloud \citep[LMC, beyond $\sim$ 4$\degr$ 
from its centre,][]{betal98} have long caught
the astronomers' interest because of the common thought that they could be old and hence, it
would be feasible from them to reconstruct the early galaxy formation and chemical enrichment
history. Indeed, the spatial distribution of the studied star clusters shows that the outer 
disc is mainly populated by those of intermediate-age ($\ga$ 2 Gyr) and old ones as well, in contrast with the much more numerous and younger
star clusters that populate the inner disc \citep{getal10,piattietal2009}.
From a chemical evolution point of view, the outer disc is commonly featured as a more metal-poor 
structure ([Fe/H] $\la$ -0.5 dex) than the inner LMC body \citep{hz09,meschin14}.

Recently, a network of streams surrounding the LMC have been discovered \citep[see, e.g.][]{bk2016}. They could be ram-pressure tails and relics of the collision between both Magellanic Clouds (MCs) 
\citep{hammeretal2015,salemetal2015} 
and hence they could contain young star clusters. Indeed, the Magellanic Bridge harbours very young and intermediate-age
star clusters as a result of the in situ star formation and stripping from tidal interaction
between both galaxies \citep{bicaetal2015}.

In this Letter we introduce a new star cluster, discovered towards the
Eastern part of the LMC outer disc and located at the Small Magellanic Cloud (SMC) distance. Its
 relatively low surface brightness could make it undetectable by previous LMC star cluster
 cataloguing  efforts. The new star cluster is unexpectedly a relatively young object with a slighlty subsolar global metal content. In order to unveil its origin, we took into account all the
star cluster properties derived here, and from them we speculate with the
possibility of being first discovered star cluster that was born in the LMC and soon ejected into 
the intergalactic space during the recent Milky Way/MCs interaction \citep{kallivayaliletal2013,casettidinescuetal2014,is2015}.

\section{star cluster discovery and its fundamental parameters}

We searched for loose star clusters in the outer regions of the L/SMC 
by using Washington $CT_1$ images obtained at the Cerro Tololo 
Inter-American Observatory 4-m Blanco telescope with the Mosaic II camera attached 
(a 8K$\times$8K CCD detector array, 36$\times$36 arcmin$^2$), that are available at the 
National Optical Astronomy Observatory (NOAO) Science Data Management 
Archives\footnote{http://www.noao.edu/sdm/archives.php.}.
The images were reduced and the photometric catalogues produced by \citet{p12a}, \citet{pietal12} 
and \citet{piatti15}, respectively. The 30 fields surveyed amounts a total
 area of 324$\degr$$^2$. 
The 50 per cent completeness level of the resulting photometry is located at a $T_1$ 
magnitude and a $C-T_1$ colour corresponding to the Main Sequence (MS) turnoff of a 
stellar population with an age $\ga$ 10 Gyr. 

\begin{figure*}
\includegraphics[width=\columnwidth]{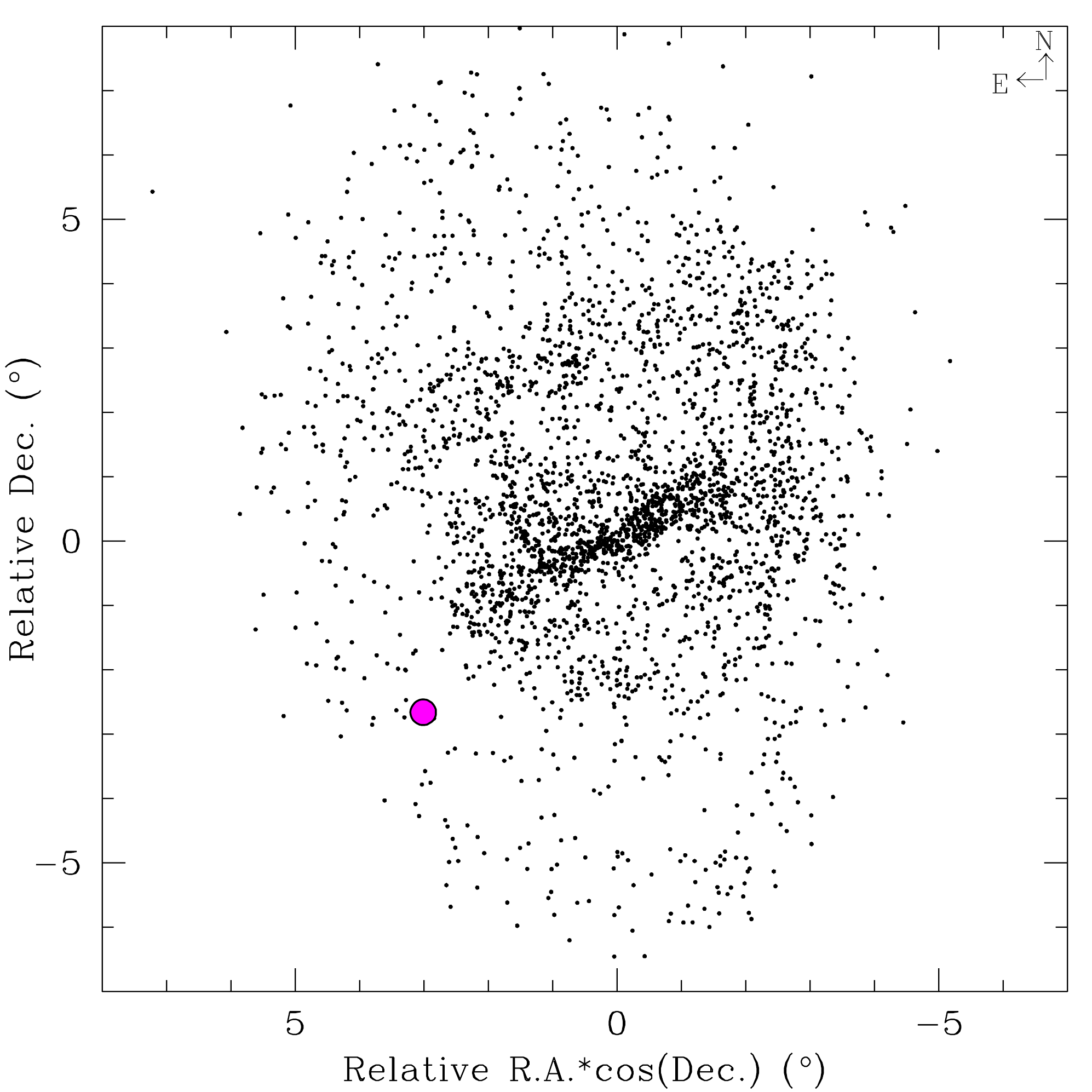}
\fbox{\includegraphics[width=\columnwidth]{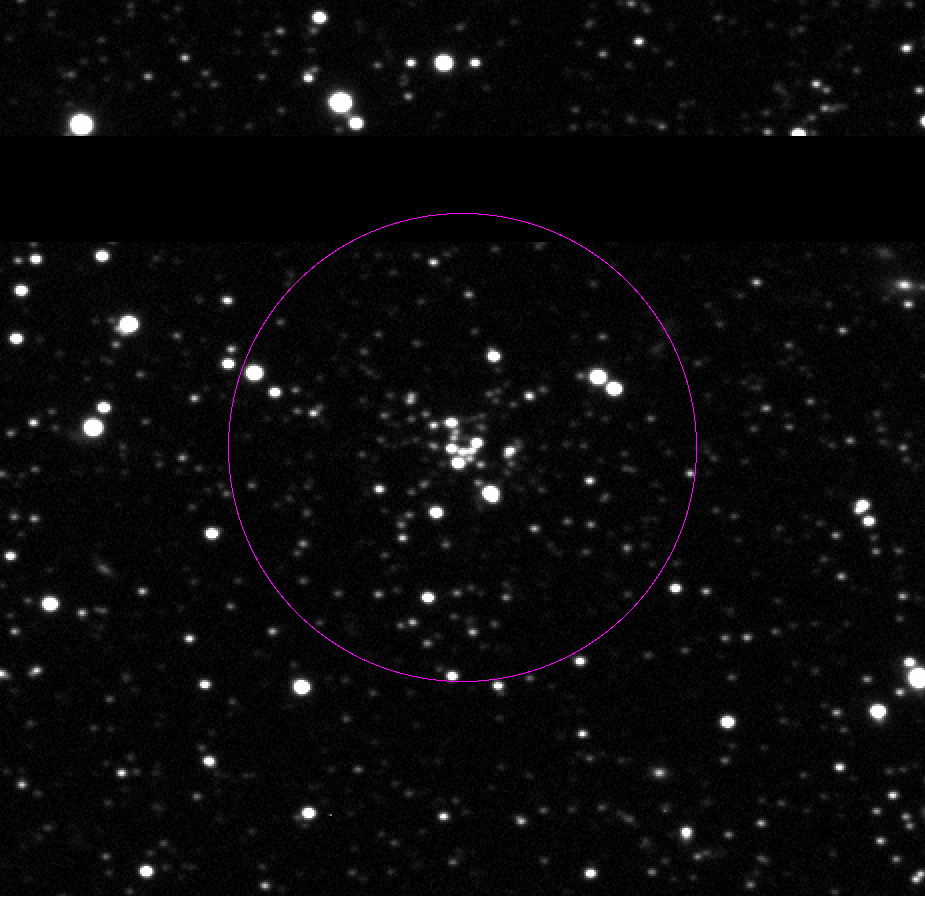}}
\caption{{\it Left:} Spatial distribution of the Bica et al. (2008)'s catalogue of star
clusters in the LMC centred at R.A. = 05$^h$ 23$^m$ 34$^s$ , Dec. = -69$\degr$ 
45$\arcmin$ 22$\arcsec$ (J2000), projected onto the sky. The discovered object is highlighted
with a magenta-coloured filled circle. {\it Right:} A 1$\arcmin$$\times$1$\arcmin$ $R$  image centred on the new LMC star cluster.  North is up and East to the left. The star cluster radius is also superimposed.}
\label{fig:fig1}
\end{figure*}

The search was performed by employing AstroML routines \citep[][and reference therein for a 
detail description of the complete AstroML package and user's Manual]{astroml} (), a machine learning
and data mining for Astronomy package. We used two different kernel density estimators 
(KDEs), namely, {\it gaussian} and {\it tophat}, and bandwidths from 0.2 up to 1.0 arcmin for
each L/SMC field photometric catalogue with stars measured in the two $CT_1$ filters.
From the total number of stellar overdensities detected per field, we imposed a cut off density 
of 3-$\sigma$ above the background level and merged the resulting lists, avoiding repeated 
findings from different runs with different bandwidths. We finally identified one new star
cluster from the L/SMC fields surveyed (see Fig.~\ref{fig:fig1}). Its central coordinates 
are listed in Table~\ref{tab:table1}. 

\begin{table}
\caption{Properties of the new star cluster.}
\label{tab:table1}
\begin{tabular}{@{}ll}\hline
Parameter & Value\\
\hline
Equatorial coords. & $\alpha$$_{J2000}$ = 6$^h$ 3$^m$ 25.46$^s$\\
                   & $\delta$$_{J2000}$ = -72$\degr$ 24$\arcmin$ 54.11$\arcsec$\\ 
Galactc coords.    & {\it l} = 283.108$\degr$\\
                   & b = -29.3795$\degr$\\
Distance modulus   & $(m-M)_o$ = 18.90 $\pm$ 0.05 mag\\
Distance           & $d$ = 60.3 $\pm$ 1.4 kpc\\
Reddening          & $E(B-V)$ = 0.08 $\pm$ 0.01 mag\\
log(Age)           & log($t$ yr$^{-1}$) = 8.45 $\pm$ 0.05\\
Age                & $t$ = 280$^{+35}_{-30}$ Myr\\
Metallicity        & [Fe/H] = -0.10 $\pm$ 0.05 dex\\
Total mass         & $M$ = 650 $\pm$ 100 $\msun$ \\
Radius             & $r$ = 9.23 $\pm$ 2.69 pc\\
Core radius        & $r_c$ = 2.76 $\pm$ 0.39 pc\\
Half-mass radius   & $r_h$ = 5.33 $\pm$ 0.51 pc\\
Tidal radius       & $r_t$ = 19.72 $\pm$ 3.94 pc\\
Jacobi radius      & $r_j$ = 13.5 $\pm$ 2.3 pc\\

\hline
\end{tabular}
\end{table}

\begin{figure*}
\includegraphics[width=\textwidth]{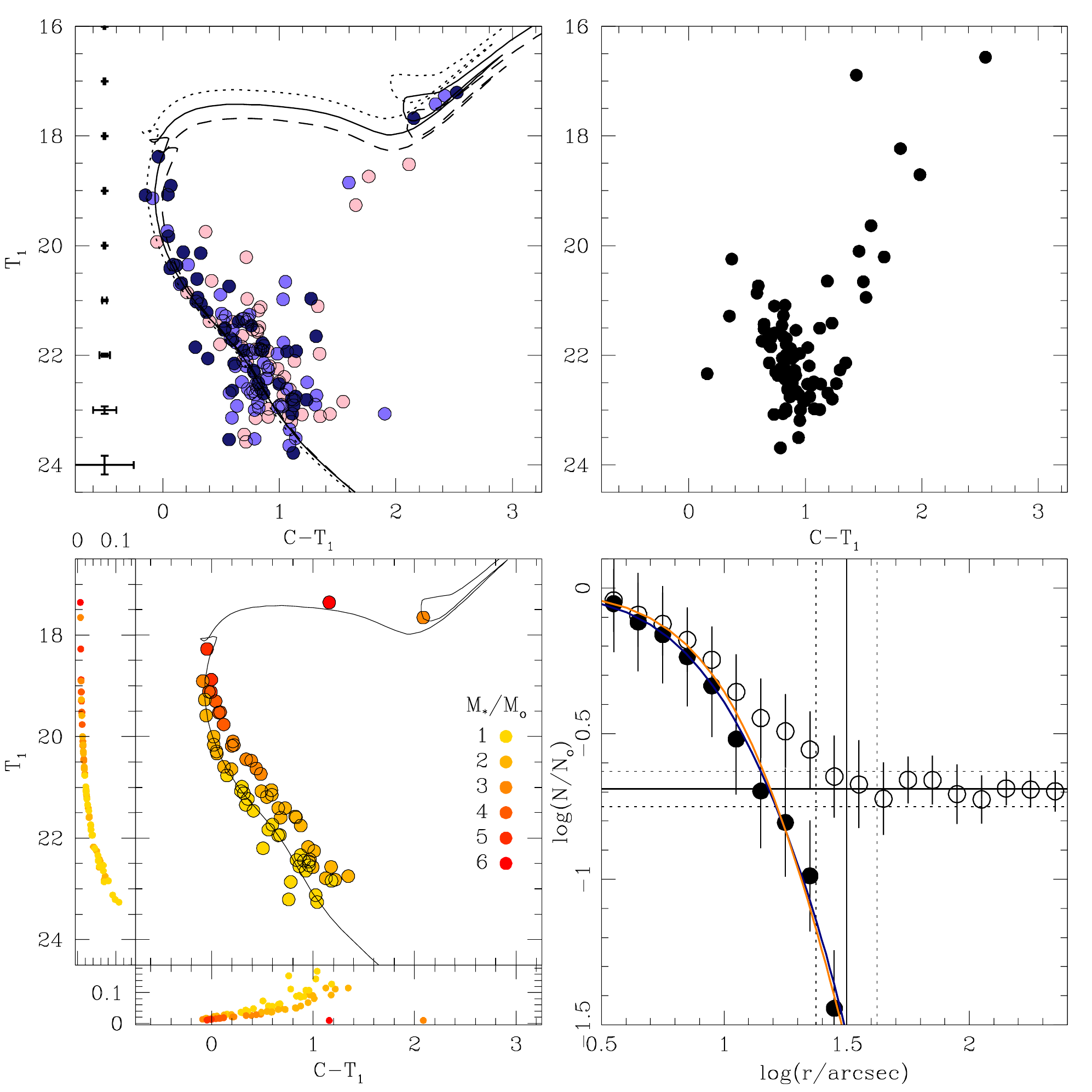}
\caption{{\it Top-left:} The cleaned star cluster CMD. Colour-scaled symbols represent stars
that statistically belong to the field (P $\le$ 25 per cent, pink), stars that might
belong to either the field or the cluster (P $=$ 50 per cent, light blue), and
stars that predominantly populate the cluster region (P $\ge$ 75 per cent, dark
blue). Three isochrones from \citet{betal12}
for log(t yr$^{-1}$) = 8.40 ($Z$ = 0.011), 8.45 ($Z$ = 0.013) and 8.50 ($Z$ = 0.014)
are also superimposed (see text for details). 
{\it Top-right: } A star field CMD for an annulus - outer and inner radii equal to 2.23 and
2.0 times the star cluster radius - centred on the star cluster.
{\it Bottom-left:} Best-generated star cluster CMD with the uncertainties in $T_1$ and
$C-T_1$, the stellar masses in colour-scaled filled circles, and the best-fitted
theoretical isochrone superimposed.
{\it Bottom-right:} Density profile obtained from star counts. Open and filled circles 
refer to measured and
background subtracted density profiles, respectively. Blue and orange solid lines
depict the fitted King and Plummer curves, respectively.}
\label{fig:fig2}
\end{figure*}

We built the star cluster density profile based on completeness corrected star counts previously performed within boxes of 5 up to 30 pixels a side distributed 
throughout the whole field of the star cluster. The selected size range of the boxes allowed us to sample statistically the stellar spatial distribution. 
Thus, the number of stars per unit area at a given radius, $r$, can be directly calculated through 
the expression:
\begin{equation}
(n_{r+} - n_{r-b/2})/(m_{r+b/2} - m_{r-b/2}),
\end{equation}

\noindent where $n_j$ and $m_j$ represent the number of stars and boxes included in a circle of radius $j$, and $b$ the box size,
respectively. Note that this method does not necessarily require a complete circle of radius $r$ within the observed 
field to estimate the mean stellar density at that distance. We used eq. (1)  because of 
the horizontal image gap (see Fig.~\ref{fig:fig1}) and the need of having a stellar 
density profile which extends far away from the star cluster centre to estimate the background 
level with high precision. This is necessary to derive the cluster radius (see 
Table~\ref{tab:table1}). The resulting mean 
density profile is shown in  Fig.~\ref{fig:fig2} (bottom-right panel).
In the figure, we 
represent the constructed and
background subtracted density profiles with open and filled circles, respectively.
Errorbars represent rms errors, to which we added the mean error of the
background star count to the background subtracted density profile. The background level
and the cluster radius are indicated by solid horizontal
and vertical lines, respectively; their uncertainties are in dotted lines.

The background corrected density profile was fitted using a \citet{king62}'s model through the expression :
\begin{equation}
 N \varpropto ({\frac{1}{\sqrt{1+(r/r_c)^2}} - \frac{1}{\sqrt{1 + (r_t/r_c)^2}}})^2
\end{equation}

\noindent where 
%$A$ is a constant, and 
$r_c$ and $r_t$ are the core and tidal radii, respectively (see Table~\ref{tab:table1} and Fig.~\ref{fig:fig2}). 
%The values derived for $r_c$ and 
%$r_t$ from the fit are listed in Table~\ref{tab:table1}, while the respective King's curve is plotted with a blue solid line in Fig.~\ref{fig:fig2}.
%and \ref{fig:fig8}. 
As can be seen, the King profile satisfactorily reproduces the whole cluster extension.
Nevertheless, in order to get independent estimates of the star cluster half-mass radius, we
%regions and slightly underestimate the cluster extensions, a behaviour found in clusters not tidally truncated
%\citep{miocchietal13,kimetal15,dalessandroetal15}. In order to account for such extra-tidal stars we 
fitted a Plummer's profile using the expression:
\begin{equation}
N \varpropto \frac{1}{(1+(r/a)^2)^2} 
\end{equation}

\noindent where $a$ is the Plummer's radius, which is related to the half-mass radius ($r_h$) by the relation $r_h$ $\sim$ 1.3$a$ (see Table~\ref{tab:table1} and Fig.~\ref{fig:fig2}). 
%The resulting $r_h$ value is listed in Table~\ref{tab:table1} and the corresponding Plummer's curve drawn with an orange solid line in Fig.~\ref{fig:fig2}.

In order to clean the star cluster colour-magnitude diagram (CMD) from the unavoidable field star contamination 
we applied a procedure developed by \citet{pb12}. 
%which has proved to be useful in previous papers, among them, \citet{p12b,p14,petal15b}.
In short, the star field cleaning relies on the comparison of each
of four previously defined field CMDs to the cluster CMD and subtracted from the latter
a representative field CMD in terms of stellar density, luminosity function, and colour
distribution. This was done by comparing the numbers of stars counted in boxes distributed 
in a similar manner throughout all CMDs. The boxes were allowed to vary in size
and position throughout the CMDs in order to meaningfully represent the actual distribution 
of field stars. Since we repeated this task for each of the four field CMDs, we could assign a membership 
probability to each star in the cluster CMD. This was done by counting the number of times
a star remained unsubtracted in the four cleaned cluster CMDs and by subsequently dividing 
this number by 4. 
Thus, we distinguished field populations projected on to the star cluster area, 
i.e. those stars with a probability $P \le$ 25 per cent, stars that could equally
likely be associated with either the field or the object of interest ($P =$ 50 per cent), 
and stars that are predominantly found in the cleaned star cluster CMDs ($P \ge$ 75 per cent) rather 
than in the field star CMDs. 
We employed this field star decontamination procedure to 
clean a circular area of radius three times that of the star cluster around its central coordinates.
Fig.~\ref{fig:fig2} (top-left panel) shows the resulting cleaned CMD for stars located 
within the star cluster radius. As can be seen, the distribution of stars with $P \ge$ 75 per cent 
resembles that of a relatively young star cluster.
For comparison purposes
we show in the top-right panel a field star CMD using an area
placed in a ring with outer and inner radii of 2.23 and 2.0 times the star cluster radius.

In order to derive star cluster' astrophysical properties, we employed the \texttt{ASteCA} 
suit of functions \citep{petal15} to generate synthetic CMDs
of star clusters covering ages from log($t$ yr$^{-1}$) = 8.0 up to 9.0 
($\Delta$log($t$ yr$^{-1}$) = 0.05),  metallicities in the range $Z$ = 0.003 - 0.025 
($\Delta$$Z$ = 0.001), interstellar extinction between 0.0 and 0.3 mag 
($\Delta$$E(B-V)$ = 0.01 mag), distance modulus between 18.0 and 19.5 mag  
($\Delta$$(m-M)_o$ = 0.05 mag) and total mass in the range 100 - 1000 $\msun$ 
($\Delta$$M$ = 50$\msun$), respectively. In total, we used $\approx$ 8.8$\times$10$^6$ models. 
%\texttt{ASteCA} was designed as a set of open source tools for an
%objective and automatic analysis of large cluster data sets. 
%The code performs a synthetic cluster-based best isochrone matching method to 
%simultaneously estimate the star cluster' properties without any biases or new correlations
%between the various derived star cluster parameter values.

The steps by which a synthetic star cluster for a given set of age, metalicity, distance modulus, and reddening values is generated by \texttt{ASteCA} is as follows: i) a theoretical isochrone
is picked up, densely interpolated to contain a thousand points throughout its entire length,
including the most evolved stellar phases. ii) The isochrone is shifted in colour and
magnitude according to the $E(B-V)$ and $(m-M)_o$ values to emulate the effects these extrinsic 
parameters have over the isochrone in the CMD. 
%At this stage the synthetic star cluster 
%can be objectively identified as a unique point in the 4-dimensional space of 
%parameters ($E(B-V)$, $(m-M)_o$, age and metallicity). 
iii) The isochrone is trimmed down to a certain faintest magnitude 
according to the limiting magnitude thought to be reached. iv) An initial mass function 
(IMF) is sampled in the mass range $[{\sim}0.01{-}100]\,M_{\odot}$ up
to a total mass value $M$ provided that
ensures the evolved CMD regions result properly populated.
%Currently, \texttt{ASteCA} lets the user choose 
%between three IMFs \citep{Kroupa_1993,Chabrier_2001,Kroupa_2002} but there is 
%no limit in the number of distinct IMFs that can be added. 
The distribution of masses is then used to obtain a properly populated synthetic 
star cluster by keeping one star in the interpolated 
isochrone for each mass value in the distribution. v) A random fraction of stars are 
assumed to be binaries, which is set by default to  
$50\%$ \citep{von_Hippel_2005}, with secondary masses 
drawn from a uniform distribution between the mass of the primary star and a 
fraction of it given by a mass ratio parameter set to $0.7$. 
%Both quantities can be modified through the input data file. 
vi) An appropriate
magnitude completeness and an exponential photometric error functions are
finally applied to the synthetic star cluster. 

As for our purposes, we used the theoretical isochrones computed by \citet{betal12}
using extensive tabulations of bolometric corrections with uncertainties $\sim$ 0.001 mag
for the $C$ and $T_1$ filters and the
IMF of \citet{kroupa02}. Fig.~\ref{fig:fig2} shows with a solid line 
the best-fitted theoretical isochrone to stars with $P \ge$ 50 per cent, which corresponds to maximum likelihood values of:
$E(B-V)$ = 0.08 mag, $(m-M)_o$ = 18.9 mag, log($t$ yr$^{-1}$) = 8.45 and $Z$ = 0.013,
respectively. In order to visually check the parameter dispersion, we bracketed that
isochrone with two ones for the following parameter values: $E(B-V)$ = 0.07 mag, $(m-M)_o$ = 18.85
mag, log($t$ yr$^{-1}$) = 8.40 and $Z$ = 0.011 (dotted line), and 
$E(B-V)$ = 0.08 mag, $(m-M)_o$ = 18.95
mag, log($t$ yr$^{-1}$) = 8.50 and $Z$ = 0.014 (short-dashed line), respectively. 
The best synthetic star cluster CMD is depicted in the bottom-left panel of the figure, with
the generated uncertainties in $T_1$ and $C-T_1$, the range of stellar masses drawn in
colour-scaled filled circles and the
best-fitted theoretical isochrones superimposed. 
The resulting mean values and errors for the different star 
cluster's properties are listed in Table~\ref{tab:table1}.

From the derived mass we estimated both the Jacobi tidal radius and the half-mass relaxion time of 
the star cluster. The former was computed from the expression \citep{cw90}:
\newpage
\begin{equation}
r_J = (\frac{M_{cls}}{3 M_{gal}})^{1/3}\times d_{GC}
\end{equation}

\noindent where $M_{cls}$ is the total star cluster mass, $M_{gal}$ is the LMC mass 
inside 8.7 kpc ((1.7$\pm$0.7)$\times$10$^{10}$ M$_{\odot}$,\citet{vdmk14}), and $d_{GC}$ is the star cluster
deprojected galactocentric distance (4.532$\degr$). The resulting Jacobi radius compares well
within the errors with the star cluster tidal radius, which suggests that the
star cluster is not tidally truncated, i.e., it is not limited.
This means that the star cluster is not expected to have lost significant amounts of stellar 
mass, so that its current mass should reflect its initial mass. Additionally, we found a half-mass density of 1.0 M$_{\odot}$ pc$^3$. This value is much larger than the minimum density a star cluster needs to have in order 
to be stable against the tidal disruption of a galaxy ($\sim$ 0.1M$_{\odot}$ pc$^{3}$, \citet{bok34}).
Accordingly, \citet{wilkinsonetal03} also showed that the tidal field of the LMC does not cause any 
perturbation on the clusters.

On the other hand, for the half-mass relaxation times we used the equation \citep{sh71}:
\begin{equation}
t_r = \frac{8.9\times 10^5 M_{cls}^{1/2} r_h^{3/2}}{\bar{m} log_{10}(0.4M_{cls}/\bar{m})}
\end{equation}

\noindent where $M_{cls}$ is the cluster mass, $r_h$ is the half-mass
radius and $\bar{m}$ is the average mass of the star cluster stars (2.6$\pm$1.2$\msun$ from
the generated synthetic CMD). The derived relaxation time resulted to be  $t_r$ = 53$\pm$15 Myr.
 
\section{discussion and conclusion}

The derived $E(B-V)$ colour excess is in excellent agreement with the values obtained from
both \citet[][0.06 mag]{hetal11} and \citet[][0.09 mag]{sf11} extinction maps, respectively, so that we infer 
that the star cluster is neither projected behind of, nor embedded into dense clouds of gas 
and dust. The star cluster is located at a distance of 60.3 kpc from the Sun, 11.3 kpc from 
the LMC centre and at a LMC angular distance of 4.532$\degr$ East. Its spatial position
and low reddening suggest that there is no dense streams behind the LMC in the star cluster
line-of-sight,  although streams have recently appeared to be more commmon around the LMC
\citep{hammeretal2015,salemetal2015}. If our derived star cluster distance were wrong,
i.e, the object should belong to the LMC disc according to its projected position in the sky, 
then it would be expected to have an age
similar to those star clusters belonging to the outer LMC disc. However,  
three star clusters with age estimate, out of five star clusters located within a radius 
of 0.75$\degr$ in the sky around the new star cluster are much older, log($t$ yr$^{-1}$) = 9.25,
which is the age asssociated to the LMC outer disc \citep{piattietal2009}.

The derived age and metallicity agrees well with the global age-metallicity relationship (AMR)
obtained by \citet[][see their Fig. 6]{pg13} for field stars and star clusters, respectively.
The AMR of the SMC --considering either field stars or star clusters - follows a clearer different
trend. It is in general $\sim$ 0.4 dex more metal-poor in [Fe/H] than that of the LMC
 \citep{p11b,petal15a}. This
could imply that the new star cluster hardly possible was born in the SMC (present star 
cluster-SMC distance of 26.4 kpc) and then stripped
by the LMC. It would be less conflicted to speculate with the possibility that
it has been born in the LMC. Indeed, besides having an age and a metallicity compatible with
the LMC AMR, there is also a good agreement for its age and mass with the age versus  mass
relationship shown by \citet{baetal13}. In addition, its structural parameters ($r_c$, $r$, $r_t$) 
are all within the frequent values found for LMC star clusters \citep{wz11}.

An unavoidable question arises: How to explain the presence of a star cluster located at the
SMC distance from the Sun and 26.4 kpc far away from the SMC centre with astrophysical properties 
(metallicity, mass, structural parameters) which resemble those of relatively young LMC star clusters? We think that the star cluster could have recently been ejected from the LMC inner body
as a consequence of tidal interaction with the Milky Way/SMC. Indeed, close Milky Way/SM passages 
have been predicted from computation of their orbital motions \citep{kallivayaliletal2013}. 
Nevertheless, spectroscopic observations for radial velocity and 
chemical abundance measurements are needed to further understand its origin. 
Furthermore, in order to constrain the effectiveness of star cluster scartering during
galaxy interaction, it would be worth to search for additional star clusters
from, for instance, the DECam survey of the MCs \citep{nideveretal2013}.

\section*{Acknowledgements}
We thank the anonymous referee whose comments and suggestions
allowed us to improve the manuscript.

%%%%%%%%%%%%%%%%%%%%%%%%%%%%%%%%%%%%%%%%%%%%%%%%%%



\bibliographystyle{mnras}
%\bibliography{paper} % if your bibtex file is called paper.bib

%to be uncommented before sending to editor
\input{paper.bbl}

%\begin{thebibliography}{99}
%\end{thebibliography}

%%%%%%%%%%%%%%%%%%%%%%%%%%%%%%%%%%%%%%%%%%%%%%%%%%

%%%%%%%%%%%%%%%%% APPENDICES %%%%%%%%%%%%%%%%%%%%%

%\appendix



%%%%%%%%%%%%%%%%%%%%%%%%%%%%%%%%%%%%%%%%%%%%%%%%%%

% Don't change these lines
\bsp	% typesetting comment
\label{lastpage}
\end{document}